\documentclass[dvips]{ws-ijmpd}

\renewcommand\ArtDir{./}

\begin{document}

\def\nocropmarks{\vskip5pt\phantom{cropmarks}}

\let\trimmarks\nocropmarks      

\markboth{R. Ruffini, C.L. Bianco, P. Chardonnet, F. Fraschetti, V. Gurzadyan, S.-S. Xue}{Emergence of a filamentary structure in the fireball from GRB spectra}

\catchline{}{}{}

\title{EMERGENCE OF A FILAMENTARY STRUCTURE IN THE FIREBALL FROM GRB SPECTRA}

\author{\footnotesize REMO RUFFINI, CARLO LUCIANO BIANCO and SHE-SHENG XUE}
\address{ICRA --- International Center for Relativistic Astrophysics and Dipartimento di Fisica,\\ Universit\`a di Roma ``La Sapienza'', Piazzale Aldo Moro 5, I-00185 Roma, Italy.}

\author{\footnotesize PASCAL CHARDONNET}
\address{ICRA --- International Center for Relativistic Astrophysics and Universit\'e de Savoie,\\ LAPTH - LAPP, BP 110, F­74941 Annecy-le-Vieux Cedex, France.}

\author{\footnotesize FEDERICO FRASCHETTI}
\address{ICRA --- International Center for Relativistic Astrophysics and Universit\`a di Trento,\\ Via Sommarive 14, I-38050 Povo (Trento), Italy.}
 
\author{\footnotesize VAHE GURZADYAN}
\address{ICRA --- International Center for Relativistic Astrophysics and Yerevan Physics Institute,\\ Alikhanian Brothers Street 2, 375036, Yerevan-36, Armenia.}

\maketitle

\pub{Received (received date)}{Revised (revised date)}

\begin{abstract}
It is shown that the concept of a fireball with a definite filamentary structure naturally emerges from the analysis of the spectra of Gamma-Ray Bursts (GRBs). These results, made possible by the recently obtained analytic expressions of the equitemporal surfaces in the GRB afterglow, depend crucially on the single parameter ${\cal R}$ describing the effective area of the fireball emitting the X- and gamma ray radiation. The X- and gamma ray components of the afterglow radiation are shown to have a thermal spectrum in the co-moving frame of the fireball and originate from a stable shock front described self-consistently by the Rankine-Hugoniot equations. Precise predictions are presented on a correlations between spectral changes and intensity variations in the prompt radiation verifiable, e.g., by the Swift and future missions. The highly variable optical and radio emission depends instead on the parameters of the surrounding medium. The GRB 991216 is used as a prototype for this model.
\end{abstract}


\section{Introduction}

It is generally agreed that the GRB afterglow originates from an ultrarelativistic shell of baryons with an initial Lorentz factor $\gamma_\circ\sim 200$--$300$ with respect to the interstellar medium (see e.g. Ruffini et al.\cite{Brasile}, Bianco \& Ruffini\cite{EQTS_ApJL} and references therein). Using GRB991216 as a prototype, in Ruffini et al.\cite{lett1,lett2} we have shown how from the time varying bolometric intensity of the afterglow it is possible to infer the average density $\left<n_{ism}\right>=1$ particle/cm$^3$ of the InterStellar Medium (ISM) in a region of approximately $10^{17}$ cm surrounding the black hole originating the GRB phenomenon. It was shown in Ruffini et al.\cite{rbcfx02} that the theoretical interpretation of the intensity variations in the prompt phase in the afterglow implies the presence in the ISM of inhomogeneities of typical scale $10^{15}$ cm. Such inhomogeneities were there represented for simplicity as spherically symmetric over-dense regions with $\left<n_{ism}^{od}\right> \simeq 10^2\left<n_{ism}\right>$ separated by under-dense regions with $\left<n_{ism}^{ud}\right> \simeq 10^{-2}\left<n_{ism}\right>$ also of typical scale $\sim 10^{15}$ cm in order to keep $\left<n_{ism}\right>$ constant. The goal of the present letter is to show how the GRB spectral observations allows to reconstruct the full 3-dimensional ISM filamentary structure with $\left<n_{ism}^{fil}\right> \simeq {\cal R}^{-1} \left<n_{ism}^{od}\right>$, where the factor ${\cal R}$ describes the effective area of the fireball emitting the X and gamma radiation.

The main results are:\\
{\bf a)} The X and gamma radiation in the afterglow has a sharp and stable structure with evolution toward the hard and the soft part of the spectrum describable by the Rankine-Hugoniot equations. Such radiation has a thermal spectrum in the co-moving frame of the fireball and originate from the shock front created ahead of the ultrarelativistic shell of baryons generating the afterglow.\\
{\bf b)} Necessary condition for such a process to occur is a filamentary structure in the ISM in order to explain the observed luminosity and to fulfill the necessary opacity condition.\\
{\bf c)} The knowledge of the intensity variation of the prompt radiation (Ruffini et al.\cite{rbcfx02}), of the luminosity in fixed energy bands in the afterglow (Ruffini et al.\cite{Spectr1}) and of the variability of the spectra (presented here) are then sufficient to determine self-consistently the 3-dimensional distribution of the ISM filamentary structure.\\
As in previous articles, we use GRB 991216 as a prototype for our model.

\section{The structure of the shock front}

The interaction between the ultrarelativistic shell of baryons and the ISM into which it is expanding is mediated by a shock front with three well-defined layers (see e.g. secs. 85--89, 135 of Landau \& Lif{\v s}its\cite{ll}, ch. 2 and sec. 13--15 of Zel'dovich \& Rayzer\cite{zr66} and sec. IV, 11--13 of Sedov\cite{sedov}). From the back end to the leading edge of this shock front there is:\\
{\bf a)} A compressed high-temperature layer, of thickness $\Delta'$, in front of the relativistic baryonic shell, created by the accumulated material swept up in the ISM.\\
{\bf b)} A thin shock front, with a jump $\Delta T$ in the temperature which in the comoving frame can be described by the Rankine-Hugoniot adiabatic equations:
\begin{equation}
\Delta T \simeq (3/16) m_p\delta v^2/k \simeq 1.5\times 10^{11} \left[\delta v/(10^5 km s^{-1})\right]^2 K\,,
\label{EqT}
\end{equation}
where $\delta v$ is the velocity jump, $m_p$ is the proton mass and $k$ is Boltzmann's constant.\\
{\bf c)} A pre-shock layer of ISM swept-up matter at much lower density and temperature, both of which change abruptly at the thin shock front behind it.

At larger distances ahead of the expanding fireball the ISM is at still smaller densities. The upper limit to the temperature jump at the thin shock front, given in Eq.(\ref{EqT}), is due to the transformation of kinetic energy to thermal energy, since the particle mean free path is assumed to be less then the thickness of the layer (a). The thermal emission of the observed X- and gamma ray radiation, which as seen from the observations reveals a high level of stability, is emitted in the above region (a) due to the sharp temperature gradient at the thin shock front described in the above region (b).

The optical and radio emission comes from the extended region (c). The description of such 
a region, unlike the sharp and well defined temperature gradient occurring in region (b), requires magnetohydrodynamic simulations of the evolution of the electron energy distribution of the synchrotron emission. Such analysis has been performed using 3-D Eulerian MHD codes for the particle acceleration models to produce the energy spectrum of cosmic rays at supernovae envelope fronts (see e.g. McKee \& Cowie\cite{mc75}, Tenorio-Tagle et al.\cite{tt91}, Stone \& Norman\cite{sn92}, Jun \& Jones\cite{j99}). Other challenges are the magnetic field and the instabilities. We mention two key phenomena: first, the importance of the development of Kelvin-Helmholtz and Rayleigh-Taylor instabilities ahead of the thin shock front. The second is the dual effect that the shock front has on the ISM initial magnetic field, first through the compression of the swept-up matter containing the field and secondly the amplification of the radial magnetic field component due to the Rayleigh-Taylor instability. Simulations of both effects (see e.g. Jun \& Jones\cite{j99} and references therein), modeling the synchrotron radio emission for an expanding supernova shell at various initial magnetic field and ISM parameters, shows for example that the presence of an initial tangential magnetic field component may essentially affect the resulting magnetic field configuration and hence the outgoing radio flux and spectrum. Among the additional effects to be taken into account are the initial inhomogeneity of ISM and the contribution of magnetohydrodynamic turbulence.

In the following we focus uniquely on the X- and gamma ray radiation, which appears to be conceptually much simpler than the optical and radio emission. It is perfectly predictable by a set of constitutive equations, which leads to directly verifiable and very stable features in the spectral distribution of the observed GRB afterglows. In line with the observations of GRB 991216 and other GRB sources, we assume in the following that the X- and gamma ray luminosity represents approximately 90\% of the energy flux of the afterglow, while the optical and radio emission represents only the remaining 10\%.

\section{The theoretically predicted spectra of GRB 991216}

\begin{figure}
\includegraphics[width=\hsize,clip]{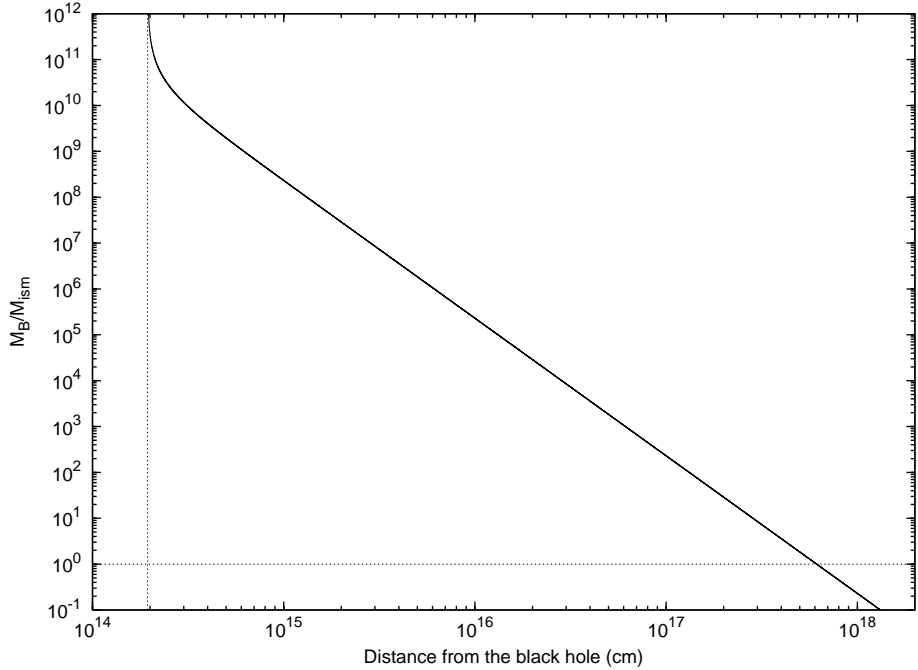}
\caption{The ratio $M_B/M_{ism}$ is plotted as a function of the radial coordinate of the expanding baryonic shell.}
\label{MBMism}
\end{figure}

For GRB 991216 the initial conditions at the beginning of the afterglow era are: $r_\circ = 1.9428 \times 10^{14}$ cm, $\gamma_\circ \simeq 310.13$, the baryonic matter $M_B \simeq 1.6122\times 10^{30}$ g, $E_{tot} \simeq 4.8298\times 10^{53}$ erg (Ruffini et al.\cite{Brasile}). By fitting the relative intensity of the Proper GRB (P-GRB, see Ruffini et al.\cite{lett2}) and of the afterglow (Ruffini et al.\cite{Brasile}), the average density has been shown to be slowly varying: $\left<n_{ism}\right> \simeq 1$ particle/cm$^3$ ($\left<n_{ism}\right> \simeq 3$ particles/cm$^3$) in the early (late) afterglow phases (Ruffini et al.\cite{Spectr1}). The integration of the Taub relativistic Rankine-Hugoniot equations of motion for the baryonic matter has been given in Bianco \& Ruffini\cite{EQTS_ApJL}. The $M_B$ is the ``piston'' in the language of Landau \& Lif{\v s}its\cite{ll}, Zel'dovich \& Rayzer\cite{zr66}, Sedov\cite{sedov}. To ensure the validity of the treatment, it is important to check that the ratio $M_B/M_{ism}$, where $M_{ism}=(4/3) \pi m_p n_{ism} (r^3-r_\circ^3)$, is always much bigger than $1$ in the entire range of the observed afterglow, which is indeed the case for $r < 4\times 10^{17}$ cm (see Fig. \ref{MBMism}).

The observed intensity variations in the ``prompt'' emission have been shown to originate in the over-dense and under-dense ISM anisotropies in Ruffini et al.\cite{rbcfx02}.

\begin{figure}
\includegraphics[width=\hsize,clip]{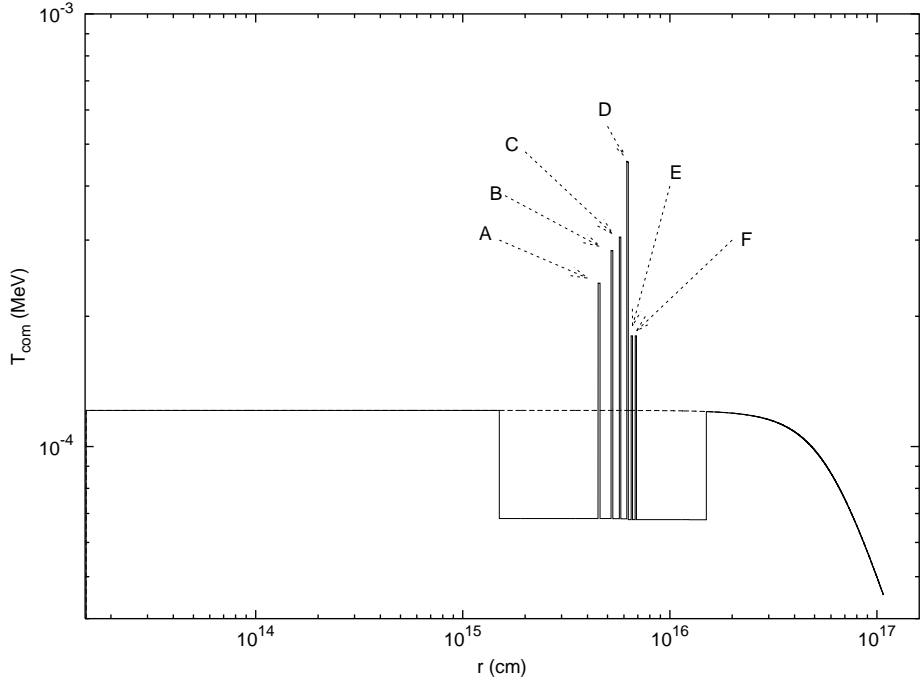}
\caption{The temperature in the co-moving frame of the shock front corresponding to the density distribution with the six spikes A,B,C,D,E,F presented in Ruffini et al.$^5$. The dashed line corresponds to an homogeneous distribution with $n_{ism}=1$.}
\label{tcom}
\end{figure}

Here the fundamental new assumption is adopted (see also Ruffini et al.\cite{Spectr1}) that the X- and gamma ray radiation during the entire afterglow phase has a thermal spectrum in the co-moving frame. The temperature is then given by:
\begin{equation}
T_s=\left[\Delta E_{\rm int}/\left(4\pi r^2 \Delta \tau \sigma {\cal R}\right)\right]^{1/4}\, ,
\label{TdiR}
\end{equation}
where $\Delta E_{\rm int}$ is the internal energy developed in the collision with the ISM in a time interval $\Delta \tau$ in the co-moving frame, $\sigma$ is the Stefan-Boltzmann constant and
\begin{equation}
{\cal R}=A_{eff}/A\, ,
\label{Rdef}
\end{equation}
is the ratio between the ``effective emitting area'' of the afterglow and the surface area of radius $r$. In GRB 991216 such a factor is observed to be decreasing during the afterglow between: $3.01\times 10^{-8} \ge {\cal R} \ge 5.01 \times 10^{-12}$ (Ruffini et al.\cite{Spectr1}).

The temperature in the comoving frame corresponding to the density distribution described in Ruffini et al.\cite{rbcfx02} is shown in Fig. \ref{tcom}.

We are now ready to evaluate the source luminosity in a given energy band. The source luminosity at a detector arrival time $t_a^d$, per unit solid angle $d\Omega$ and in the energy band $\left[\nu_1,\nu_2\right]$ is given by (see Ruffini et al.\cite{Brasile,Spectr1}):
\begin{equation}
\frac{dE_\gamma^{\left[\nu_1,\nu_2\right]}}{dt_a^d d \Omega } = \int_{EQTS} \frac{\Delta \varepsilon}{4 \pi} \; v \; \cos \vartheta \; \Lambda^{-4} \; \frac{dt}{dt_a^d} W\left(\nu_1,\nu_2,T_{arr}\right) d \Sigma\, ,
\label{fluxarrnu}
\end{equation}
where $\Delta \varepsilon=\Delta E_{int}/V$ is the energy density released in the interaction of the ABM pulse with the ISM inhomogeneities measured in the comoving frame, $\Lambda=\gamma(1-(v/c)\cos\vartheta)$ is the Doppler factor, $W\left(\nu_1,\nu_2,T_{arr}\right)$ is an ``effective weight'' required to evaluate only the contributions in the energy band $\left[\nu_1,\nu_2\right]$, $d\Sigma$ is the surface element of the EQTS at detector arrival time $t_a^d$ on which the integration is performed (see also Ruffini et al.\cite{rbcfx02}) and $T_{arr}$ is the observed temperature of the radiation emitted from $d\Sigma$:
\begin{equation}
T_{arr}=T_s/\left[\gamma \left(1-(v/c)cos\vartheta\right)\left(1+z\right)\right]\, .
\label{Tarr}
\end{equation}

The ``effective weight'' $W\left(\nu_1,\nu_2,T_{arr}\right)$ is given by the ratio of the integral over the given energy band of a Planckian distribution at a temperature $T_{arr}$ to the total integral $aT_{arr}^4$:
\begin{equation}
W\left(\nu_1,\nu_2,T_{arr}\right)=\frac{1}{aT_{arr}^4}\int_{\nu_1}^{\nu_2}\rho\left(T_{arr},\nu\right)d\left(\frac{h\nu}{c}\right)^3\, ,
\label{effweig}
\end{equation}
where $\rho\left(T_{arr},\nu\right)$ is the Planckian distribution at temperature $T_{arr}$:
\begin{equation}
\rho\left(T_{arr},\nu\right)=\left(2/h^3\right)h\nu/\left(e^{h\nu/\left(kT_{arr}\right)}-1\right)
\label{rhodef}
\end{equation}

\begin{figure*}
\includegraphics[width=\hsize,clip]{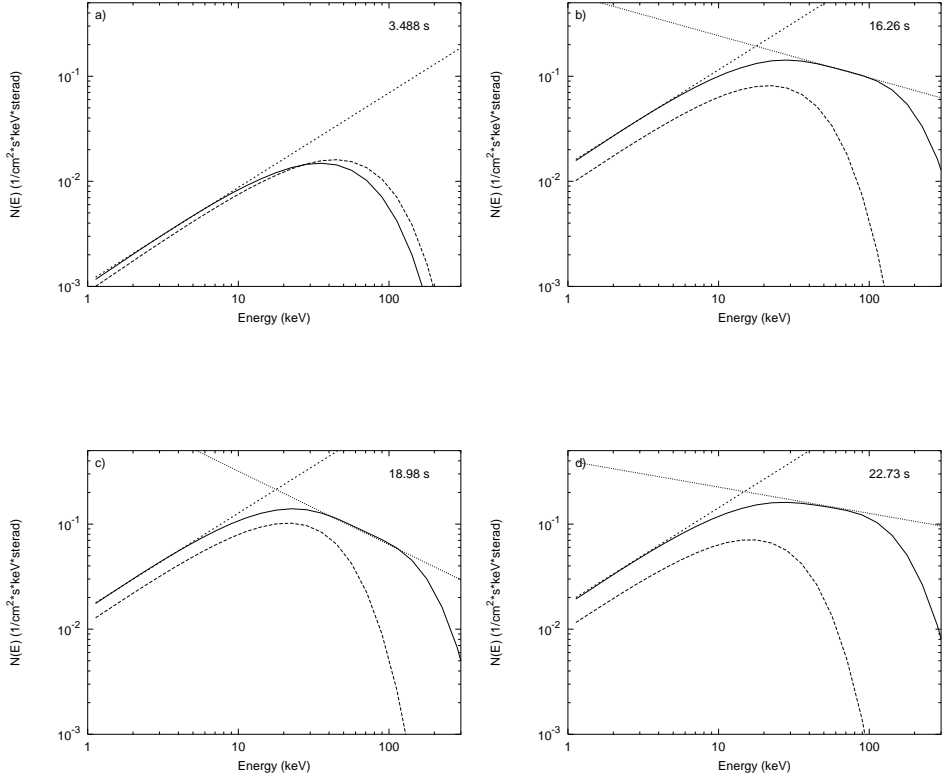}
\caption{Theoretically predicted spectra (solid curves) for selected values of the photon arrival time at the detector: $t_a^d = 3.488$ s (a), $t_a^d = 16.26$ s (b), $t_a^d = 18.98$ s (c), $t_a^d = 22.73$ s (d). The dotted straight lines corresponds to the slopes $\alpha$ in the low ($\beta$ in the high) energy bands at $1$--$5$ keV ($40$--$80$ keV). It is interesting to compare and contrast the solid curves, representing the actual spectral distribution, to the corresponding boosted black body spectra at $\vartheta=0$ (dashed curves, see text).}
\label{4spect}
\end{figure*}

In Ruffini et al.\cite{Spectr1} we applied this theoretical framework for the analysis of the observed luminosities in fixed energy bands ($50$--$300$ keV and $2$--$10$ keV) of GRB991216 over an observation time lasting $10^7$ s and we gave, as well, a self-consistent explanation of the observed hard-to-soft transition of the GRB spectra and of their variability during the GRB afterglow. We here illustrate the expected spectral distributions for selected values of the photon arrival time at the detector, during the early phases of afterglow (see Fig. \ref{4spect}). These predictions appear to be of the greatest interest in view of the forthcoming Swift\footnote{http://swift.gsfc.nasa.gov/} mission and future high-precision missions. The slopes of the four spectra appear to be constant with $\alpha = 0.9$ in the range $1$--$5$ keV, while they appear to be extremely varying in the energy range $40$--$80$ keV. We have, in fact, over an arrival time interval of a few seconds (see Fig. \ref{4spect}): $\beta = -0.4$ at $t_a^d= 16.26$ s, $\beta = -0.7$ at $t_a^d = 18.98$ s, $\beta = -0.25$ at $t_a^d = 22.73$ s. It is particularly attractive to analyze observationally the very precise correlation predicted in our model between the intensity variations during the prompt emission and the effective power-law indexes of the high-energy and low-energy parts of the spectra.

The most important feature is that, although the original emission in the co-moving frame is thermal, the final theoretically predicted spectra appear to be highly non-thermal! The final spectra, in fact, originate from a superposition of an infinite series of contributions along the EQTS depending on the angle $\vartheta$ between the emission region and the line of sight (see Ruffini et al.\cite{Spectr1}). Each contribution carries with itself a distinct thermal spectrum boosted by a specific Lorentz factor. In order to visualize the departure from a thermal spectrum, we have also shown in each figure the expected spectrum from an ideal emission process in which all the energy radiated is assumed to originate at $\vartheta = 0$ with the same Lorentz gamma factor, which is of course a perfect blue-shifted black body spectrum.

\section{On the physical origin of the ${\cal R}$ factor and of the thermal radiation}

We are now left to explain the physical origin {\bf a)} of the thermalization process and {\bf b)} of the ${\cal R}$ factor. Both these problems are solved if we assume that the ISM matter in the over-dense region is distributed in a filamentary structure in which each filament has an average density
\begin{equation}
\left<n_{ism}^{fil}\right>=\left<n_{ism}^{od}\right>/\mathcal{R}\, ,
\label{nfil}
\end{equation}
which is then varying from $\left<n_{ism}^{fil}\right> \simeq 10^{9} \left<n_{ism}\right>$ at $r \simeq 1.0\times 10^{16}$ cm up to $\left<n_{ism}^{fil}\right> \simeq 10^{13} \left<n_{ism}\right>$ at $r \simeq 10^{17}$ cm. Since the X- and gamma-ray emission only occurs in the filamentary structure, it follows from Eq.(\ref{nfil}) that the effective emitting area is indeed the one given by Eq.(\ref{Rdef}).

The filamentary structure, which has the typical length of an over-dense region $\Delta R \simeq 10^{15}$ cm, is optically thick:
\begin{equation}
\tau = \sigma_{T} \left<n_{ism}^{fil}\right> \Delta R > 1\, ,
\label{tau}
\end{equation}
where $\sigma_T$ is the Thomson cross-section. The internal temperature $T_{int}$ of the compressed layer of thickness $\Delta'$ in the above region a), measured in its co-moving frame, is given by:
\begin{equation}
T_{int}=\left\{\Delta E_{\rm int}/\left[16\pi r^2 \gamma^2 \Delta \tau \sigma {\cal R} \left(\Delta' / \Delta R\right) \right]\right\}^{1/4}\, ,
\label{TdiRV}
\end{equation}
Since we must have $T_{int} > T_s$, we then have:
\begin{equation}
\Delta' < \Delta R/\left(4\gamma^2\right)\, .
\label{Delta'}
\end{equation}
 
\section{Conclusions}

Having conceptually explained both the thermalization process and the ${\cal R}$ factor as originating from the filamentary structure, the ``inverse problem'' is now open to the solution: to infer, from the observed spectra and intensity variation, the actual three dimensional extension of the filamentary structure.

The existence of filamentary structures will indeed sharpen the description of the intensity variations presented in Ruffini et al.\cite{rbcfx02}. This will in principle allow to describe all the spikes in the observed prompt radiation, as well as the event $D$ of Ruffini et al.\cite{rbcfx02}, which were not explainable in the average spherically symmetric approximation.

\end{document}